\documentclass{CUP-JNL-DTM}%

\usepackage{graphicx}
\usepackage{multicol,multirow}
\usepackage{amsmath,amssymb,amsfonts}
\usepackage{mathrsfs}
\usepackage{amsthm}
\usepackage{rotating}
\usepackage{appendix}
\usepackage[numbers]{natbib}
\usepackage{ifpdf}
\usepackage[T1]{fontenc}
\usepackage{newtxtext}
\usepackage{newtxmath}
\usepackage{textcomp}
\usepackage{xcolor}
\usepackage{lipsum}
\usepackage[colorlinks,allcolors=blue]{hyperref}

\theoremstyle{definition}

\numberwithin{equation}{section}

\jname{Data/Math}
\articletype{ARTICLE TYPE}
\jyear{YEAR}
\begin{document}

\begin{Frontmatter}

\title[ANALYSIS OF THE INFLUENCE OF FINAL RESOLUTION ON ADC ACCURACY]{ANALYSIS OF THE INFLUENCE OF FINAL RESOLUTION ON ADC ACCURACY}

% There is no need to include ORCID IDs in your .pdf; this information is captured by the submission portal when a manuscript is submitted. 
\author[1]{Anzhelika Stakhova }

\authormark{Anzhelika Stakhova } 
%\textit{et al}.

\address[1]{\orgdiv{Department of Computerized Electrical Systems and Technologies}, \orgname{National Aviation University}, \orgaddress{\city{Kyiv}, \postcode{03058},  \country{Ukraine}}}

\authormark{Anzhelika Stakhova }



\abstract{This work is devoted to the study of the influence of quantization noise on the spectral characteristics of a digital signal and the assessment of spectrum measurement errors that arise due to the quantization noise of an analog-to-digital converter. To achieve more accurate and reliable measurements of the spectrum, an error assessment was carried out, which allows taking into account the impact of quantization noise on the spectral data. This is important for obtaining more accurate results and ensuring high-quality measurements of the spectral components of the vibration signal. In addition, further research is aimed at developing methods for estimating spectrum measurement errors taking into account other possible sources of errors and contributing to the development of compensation algorithms to reduce the impact of quantization noise.}

\end{Frontmatter}

% \section*{Impact Statement}
% Some Data journals (DAP, DCE) require an `Impact Statement' section. Comment out this section if it is not required.

% Some math journals (FLO) require a table of contents. Comment out this line if no ToC is needed.
% \localtableofcontents

% \section[This is an A Head]{This is an A head this is an A head this is an A head}
% \lipsum[1]

% \subsection{This is a B head this is a B head this is a B head this is a B head}

% \lipsum[2]

% \subsubsection{This is a C head this is a C head this is a C head this is a C head}

% \lipsum[3]

% \paragraph{This is a D head this is a D head this is a D head this is a D head}

% \lipsum[4]

\section[This is an A Head]{Introduction}
%\subsection{This is a B head this is a B head this is a B head this is a B head this is a B~head}
%\subsubsection{This is a C head this is a C head this is a C head this is a C head}
%\lipsum[4]\footnote{This is sample for footnote this is sample for footnote this is sample for footnote  this is sample for footnote this is sample for footnote.}
In the modern world, where accuracy and reliability of measurements are crucial in many fields of science and technology, measurement systems have become an essential component for obtaining quality data and analyzing various phenomena.

Modern measurement systems consist of a complex of technical tools for measurement, data collection, and processing, integrated into a unified structure that operates according to specific rules [1]. Such systems include analog-to-digital converters (ADCs), which, in turn, are capable of converting analog signals into a digital format, enabling the storage and processing of data with high accuracy and efficiency. However, ADCs are not perfect and are accompanied by certain errors. One such error is quantization error, which arises due to the limited number of bits in the digital format. Quantization refers to representing an analog signal in the form of discrete values. The smaller the number of bits in the digital format, the greater the impact of quantization error on measurement accuracy.

To improve the accuracy of measurements, it is necessary to study the influence of quantization noise on the spectral characteristics of a signal in order to develop a method for estimating measurement errors in the spectrum caused by ADC quantization noise. The estimation of these errors is important for correctly interpreting measurement results and ensuring high accuracy and reliability of the measurement system.
\section[This is an A Head]{Task statement}
Quantization error in ADC arises due to the approximate representation of an infinite set of analog signal values with limited bit resolution. When using an ADC to measure vibrations of rotating and moving parts in a technological system, the continuous input signal is divided into a finite number of discrete levels, which affects measurement accuracy.

As a result of the limited precision of the ADC's bit resolution, there is a certain maximum number of values that can be represented in the digital format. This leads to the approximation of the input analog signal to the nearest discrete value, which can introduce error. The smaller the ADC's resolution, that is, the fewer bits used to represent the signal, the greater the quantization error becomes. 

The magnitude of the ADC's quantization error at the input is determined as [2]:
\begin{equation}
\Delta x_q=y-x=D(x)\cdot q-x,
\label{eq1}
\end{equation}
where $y$  is the output signal of the ADC, referred to its input, $x$  is the value of the input signal, $q$  is the value of the ADC's least significant bit (EMR, ADC quantization step),  $D(x)$ is the value of the ADC's digital code.

The error value (2.1) over time, when the number of bits be-comes asymptotically large, is commonly referred to as quantization noise. For the instantaneous value of ADC quantization noise, the following relationship holds true [3]:
\begin{equation}
\-0,5 \cdot q \leq \Delta x_q \leq 0,5 \cdot q.
\label{eq2}
\end{equation}

A significant number of publications [2-6] have been dedicated to the study of quantization noise. However, the question of the influence of finite bit resolution on spectrum deviations has been relatively understud-ied. It is necessary to develop a methodology for calculating the measure-ment error in the spectrum caused by the finite bit resolution of the ADC. In doing so, the temporal discretization of the ADC's output signal must be taken into account.

The form of quantization noise is determined by the parameters of the input signal and the ADC. For a sinusoidal input signal, the parameters that affect the magnitude of quantization noise include the amplitude, frequency, sampling frequency, and initial phase. In the case of a polyhar-monic signal, the parameters of each spectral component also come into play.

The temporal dependence of quantization noise for a sinusoidal input signal with unit amplitude is shown in Figure 1, while the corresponding amplitude spectra of quantization noise are presented in Figure 2. As observed from the dependencies in Figure 1, quantization noise does not possess a random nature, which is supported by [1, 6, 8-9]. When the sampling frequency is a multiple of the input signal frequency (Figure 1, b), periodicity in quantization noise with a period equal to the input signal period can be observed. Otherwise, this effect is not observed (Figure 1, a).

During the simulation, the following parameters were used: initial phase - 0 rad; frequency - 50 Hz; sampling frequency - 10240 Hz (a) / 10000 Hz (b); measurement interval - 0.0201 s; ADC bit resolution - 12 bits.

Figure 2 illustrates that when there is no frequency matching between the sampling frequency and the frequency of the input signal (Figure 2a), the spectrum of the quantization noise appears uniform. When it comes to matching frequencies (Figure 2b), where the spectral components have frequencies that are multiples of the input signal frequency, an increase in amplitude values can be observed. This can be attributed to the periodicity of the quantization noise as shown in Figure 1b. Additionally, as seen in Figure 2b, the amplitude values of even harmonics of the quantization noise are zero.

Currently, most ADCs have a unipolar transfer function, which reduces the number of effective bits by half. Furthermore, the amplitude coefficient should be taken into account for distortion-free signal transformation. Finally, high-resolution ADCs (particularly those with 24 bits) may have fewer effective bits at a given sampling frequency.

Therefore, the task of estimating measurement errors in the spectrum influenced by ADC quantization noise is significant.

\begin{figure}[t]%
(a)
\FIG{\includegraphics[width=0.9\textwidth]{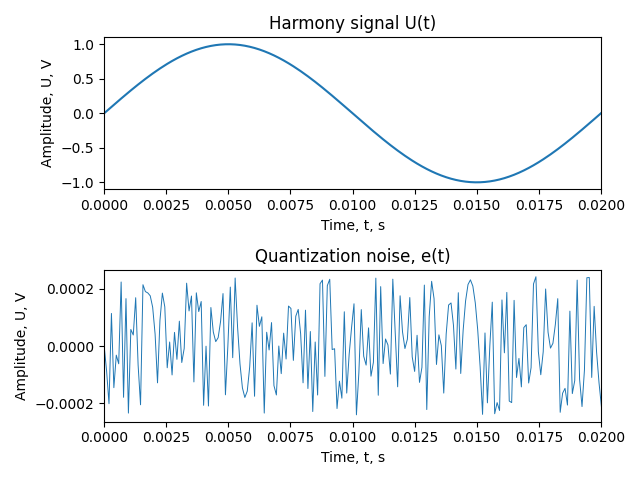}}
(b)
\FIG{\includegraphics[width=0.9\textwidth]{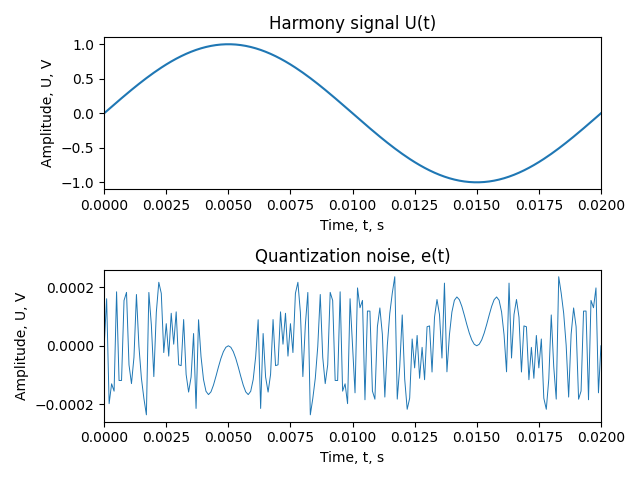}}
{\caption{The dependence of quantization error over time in the absence of frequency matching (a) and in the presence of frequency matching (b) between the sampling frequency and the input signal frequency}
\label{fig1}}
\end{figure}

\begin{figure}[t]%
(a)
\FIG{\includegraphics[width=0.9\textwidth]{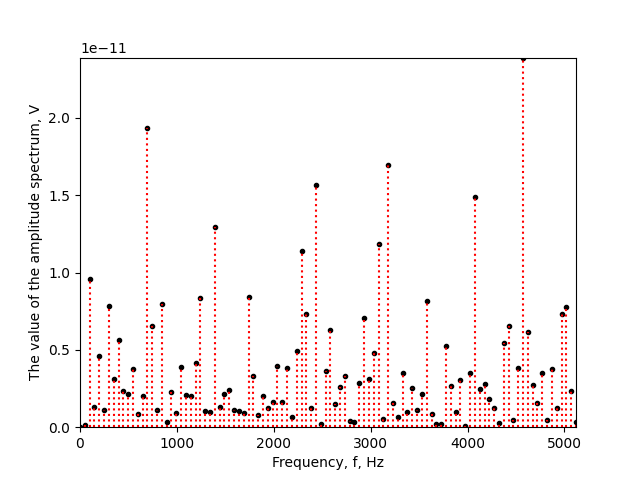}}
(b)
\FIG{\includegraphics[width=0.9\textwidth]{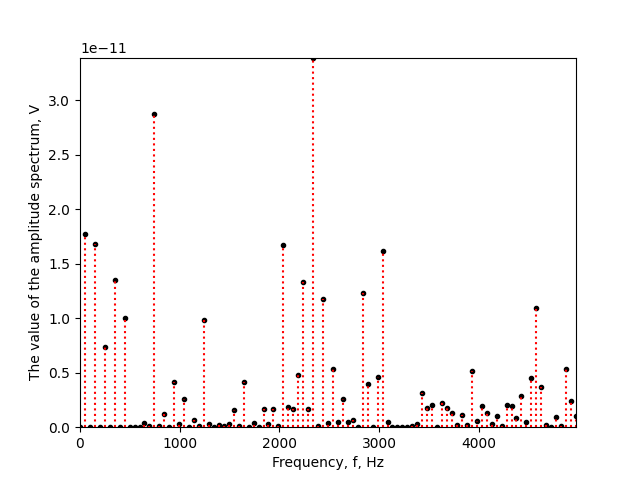}}
{\caption{The spectra of the ADC's output signal, brought to the input, under the condition of no frequency matching (a) and with frequency matching (b) between the sampling frequency and the input signal frequency}
\label{fig2}}
\end{figure}

\begin{figure}[t]%
(a)
\FIG{\includegraphics[width=0.9\textwidth]{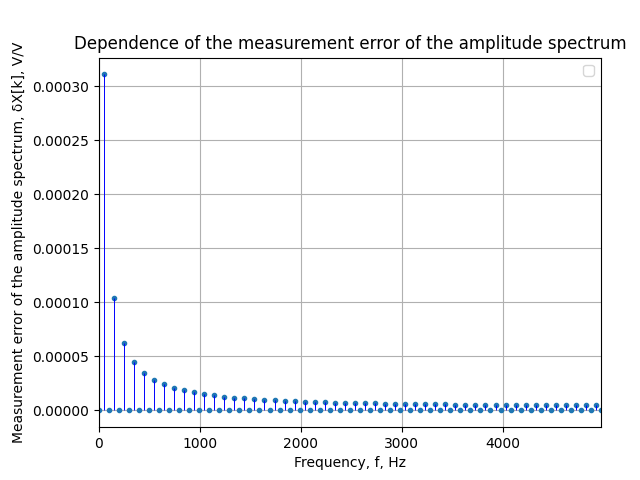}}
(b)
\FIG{\includegraphics[width=0.9\textwidth]{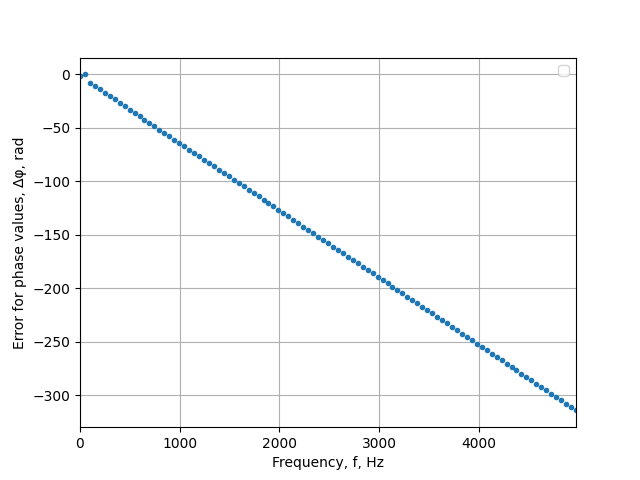}}
{\caption{The dependence of the measurement error of the amplitude (a) and phase (b) spectra for a sinusoidal input signal}
\label{fig3}}
\end{figure}

\section[This is an A Head]{Formulation of the problem}
Assessing and compensating for quantization error are important tasks in the development of measurement systems. This may involve the use of special algorithms, filters, or signal processing methods to reduce the impact of quantization error on the obtained measurement results and improve the accuracy of the measurement system.

The aim of the article is to improve the accuracy of measurement results in equipment monitoring through vibration signals. To achieve this, the methods of attaching measurement transducers in the equipment monitoring system have been investigated. These attachment methods can have an impact on the accuracy of vibration measurements and the results of equipment condition monitoring.

\section[This is an A Head]{Presentation of the material}
From the literature sources, it is known that the presence of quantization noise leads to distortion of the spectrum of the digital signal at the output of the ADC. This distortion can manifest as distorted amplitude components and additional noise components that do not exist in the original analog signal. These distortions can affect the accuracy of spectral analysis and result in inaccurate measurements or incorrect interpretation of the results.

It is known that with a sufficiently large number of ADC bits and input signal samples, it has been observed that the spectral density of quantization noise tends to adhere to a uniform distribution [2-5, 7-9]. In the Nyquist frequency band, the quantization noise spectral density can be used to calculate by the expression:
\begin{equation}
\sigma_q=q /(2 \sqrt{3}),
\label{eq3}
\end{equation}
where $\sigma_q$ is the Quantization Noise Spectral Density (QNSD) of the analyzed ADC within the Nyquist bandwidth. 

In the case of a finite number of input samples and a finite ADC reso-lution, in the scenario where there is a finite number of input samples, the relationship between the spectral density of quantization noise and its distribution deviates from a uniform distribution [2, 7-8]. Nevertheless, when the sampling frequency increases without a frequency relationship between the input signal and the sampling frequency, the shape of the quantization noise spectral density tends to approach a uniform distribu-tion.

ADC manufacturers specify a range of dynamic parameters (SNR, ENOB, SNDR, THDN) in the technical specifications of their microchips, which are related to the phenomenon of quantization noise [3-4, 10]. When calculating parameters such as SNDR and SNR, factors other than quantization noise are taken into account, such as non-linearity of the ADC transfer function [6, 10-12] and inherent ADC noise unrelated to quantization. The SNR parameter characterizes the Quantization Noise Spectral Density (QNSD) of the ADC when a sinusoidal signal is applied to its input. In addition to quantization noise, this parameter also includes the inherent noise of the ADC. Consequently, using these dynamic parameters to estimate the measurement errors of the spectrum (errors in the meas-urement values of individual spectral components) caused by quantization noise leads to a significant overestimation of these errors.

The amplitude values of vibration signal spectral components, as well as the phase shifts between components of the same frequency, are determined by performing the Discrete Fourier Transform (DFT) of the vibration signal [14-16]. The representation of the frequency domain of the ADC output signal, referenced to its input, is determined as [7-8]:
\begin{equation}
\begin{gathered}\dot{Y}[k]=\dot{X}[k]+\Delta \dot{X}[k]=\frac{2}{N} \sum_{n=0}^{N-1}\left(x[n] e^{-j \frac{2 \pi n k}{N}}\right)+\frac{2}{N} \sum_{n=0}^{N-1}\left(\Delta x_q[n] e^{-j \frac{2 \pi n k}{N}}\right)\end{gathered},
\label{eq1}
\end{equation}
where $x[n]$ - ADC input signal samples; $N$ represent the count of signal samples taken during a specific measurement interval; $\dot{Y}[k]$ -  $k$- scaled (coefficient $2 / N$ ) complex component of the spectrum of the ADC input signal; $\dot{X}[k]$ - $k$- scaled (coefficient $2 / N$ ) complex component of spectrum of the output signal of the ADC, referenced to its input; $\Delta x_q[n]$ - quantization noise samples referenced to the input of the ADC; $\Delta \dot{X}[k]$ -  $k$- scaled (coefficient $2 / N$ ) complex component of spectrum of signal $\Delta x_q$ .

In the worst case scenario, the relative measurement error of the amplitude value for any spectral component of the input signal can be obtained using the Quantization Noise Power (QNP) and is given by Equation (4.2), which is:
\begin{equation}
\begin{gathered}\left.\delta X[k]\right|_{\max }=|\delta \dot{X}[k]|_{\max }= \sqrt{2} \cdot q / \sqrt{12}=q / \sqrt{6}\end{gathered}.
\label{eq1}
\end{equation}

In other words, the Quantization Noise Power (QNP) corre-sponding to the entire frequency band is attributed to a single analyzed spectral component. This approach (Equation 4.3) will be referred to as the "Quantization Noise Power method."

Direct utilization of Equation (4.2) is complicated since the variation law of quantization noise over time for an arbitrary signal is unknown. It is possible to estimate the quantization noise values using their extreme values [17], where the quantization noise for each sample leads to an error . Additionally, analytical relationships can be used, but they are only valid for specific input signal forms [2, 7-8, 12-13].

From literary sources [17], it is known that estimates of the measurement error of Quantization Noise Power across the entire frequency band can be obtained using the extreme values of the quantization noise. In this case, it is assumed that the quantization noise takes the following form:
\begin{equation}
\begin{cases}\Delta x_q[n]=0,5 \cdot q, & x[n] \geq 0, \\ \Delta x_q[n]=-0,5 \cdot q, & x[n]<0.\end{cases}
\label{eq1}
\end{equation}

Now let's examine the implementation of approach (4.4) for estimating the measurement error of both the amplitude and phase spectrum. By substituting equation (4.4) into equation (4.2) for the scenario of a sinusoidal input signal with zero initial phase and a measurement time equal to one period ($m=1$), the resulting expression is obtained:
\begin{equation}
\begin{gathered}\Delta \dot{X}[k]=\frac{2 q \sin ^2(0,5 k \pi)}{N \sin (k \pi / N)} \left(\sin \left(\frac{\pi k(N-1)}{N}\right)+j \cos \left(\frac{\pi k(N-1)}{N}\right)\right)\end{gathered}.
\label{eq1}
\end{equation}

If the condition of sampling rate being a multiple of the input signal frequency is satisfied, for a sinusoidal input signal, the frequency domain representation can be expressed as:
\begin{equation}
\dot{X}[k]=\left\{\begin{array}{l}X_m(\sin \alpha-j \cos \alpha), \quad k=1, \\ 0, \quad k \neq 1.\end{array}\right.
\label{eq1}
\end{equation}
where  and   represents the amplitude value, and denotes the initial phase of the sinusoidal input signal to the ADC. 

In this case, the amplitude values of the output signal components of the ADC can be determined by calculating the modulus of the sum of expressions (4.5) and (4.6). It is important to note that expression (4.5) is derived when the initial phase is zero  = 0 and the measurement interval is equal to one period of the input sinusoidal signal. Thus, the resulting expression is as follows:
\begin{equation}
|\dot{Y}[k]|^2=Y^2[k] \cong \left\{\begin{array}{l}\ X_m^2-\frac{4 X_m q}{N \sin (\pi / N)}\left(\cos \left(\frac{\pi(N-1)}{N} \right)\right) , \ k=1, \\ 
\left(\frac{2 q \sin ^2(0,5 k \pi)}{N \sin (k \pi / N)}\right)^2, k \neq 1 .\end{array}\right.
\label{eq1}
\end{equation}

By using expression (4.7), the deviations of the spectral amplitude components of the vibration signal can be calculated. In expression (4.7), the form of quantization noise takes the worst-case scenario for SNR measurement error. To calculate the amplitude spectrum, it is required to take the modulus of expression (4.7). To simplify the final expression of the amplitude spectrum, the square root can be approximated by representing it as the sum of the first two terms of the Taylor series expansion. By performing these operations, we obtain:
\begin{equation}
\ Y[k] \cong \left\{\begin{array}{l}\ X_m-\frac{2 q}{N \sin (\pi / N)}\left(\cos \left(\frac{\pi(N-1)}{N}\right)\right), \ k=1, \\ \frac{2 q \sin ^2(0,5 k \pi)}{N \sin (k \pi / N)}, k \neq 1.\end{array}\right.
\label{eq1}
\end{equation}

Subsequently, the relative error of the amplitude spectrum, concerning the amplitude value of the input sinusoidal signal, induced by the ADC quantization noise (the time-dependent nature of the quantization noise as given by relation (4.4)) is determined by the following equation:
\begin{equation}
\delta X[k]=|\delta \dot{X}[k]| \cong\left\{\begin{array}{l}\ -\frac{2 q}{N X_m \sin (\pi / N)}\left(\cos \left(\frac{\pi(N-1)}{N}\right)\right), \ k=1, \\ \frac{2 q \sin ^2(0,5 k \pi)}{N X_m \sin (k \pi / N)}, k \neq 1 .\end{array}\right.
\label{eq1}
\end{equation}

Analyzing equation (4.9), it can be observed here is no error in the amplitude value for all even harmonics when the quan-tization error is applied according to the law (4.4).

To determine the phase error caused by the presence of quantization noise, it is necessary to determine the phase characteristic of equation (4.8)
\begin{equation}
\varphi_Y[k]=\operatorname{arcctg}(\operatorname{Re}(\dot{Y}[k]) / \operatorname{Im}(\dot{Y}[k])).
\label{eq1}
\end{equation}

The initial phase of the fundamental component of the output signal of the ADC, for the case when  $\alpha$ = 0 and $m=1$, can be found by substituting equations (4.5) and (4.6) into equation (4.10)
\begin{equation}
\varphi_Y[k]|_{k=1} = \operatorname{arcctg} \left ( \frac{X_m \cos \left(-\frac{\pi}{2}\right)+\frac{2 q}{N \sin (\pi / N)} \sin \left(\frac{\pi(N-1)}{N}\right)}{\ X_m \sin \left(-\frac{\pi}{2}\right)+\frac{2 q}{N \sin (\pi / N)} \cos \left(\frac{\pi(N-1)}{N}\right) } \right).
\label{eq1}
\end{equation}

Ignoring the second term in the denominator, equation (4.11) takes the form:
\begin{equation}
\varphi_Y[k]|_{k=1} = \operatorname{arcctg} \left ( \operatorname{ctg}\left(-\frac{\pi}{2}\right)-\frac{2 q \sin (\pi(N-1) / N)}{X_m N \sin (\pi / N)} \right).
\label{eq1}
\end{equation}

By expanding the obtained expression in a Taylor series up to two terms, we get:
\begin{equation}
\varphi_Y[k]|_{k=1} = -0,5\pi +\frac{2 q \sin (\pi(N-1) / N)}{X_m N \sin (\pi / N)}.
\label{eq1}
\end{equation}

Accordingly the absolute error in determining the phase spectrum for the fundamental component of the output signal of the ADC, applied to the input, is determined by the equation:
\begin{equation}
\Delta \varphi_Y[k]|_{k=1} = \frac{2 q \sin (\pi(N-1) / N)}{X_m N \sin (\pi / N)}.
\label{eq1}
\end{equation}

Let's reiterate that equation (4.14) is derived for the case of $\alpha$ = 0 and $m=1$. To determine the phase values of the harmonics in the output signal of the ADC, we need to apply equation (4.10) to equation (4.5) under the condition:
\begin{equation}
\varphi_Y[k]|_{k \neq 1} = \operatorname{arcctg} \left ( \frac{\sin (\pi k(N-1) / N)}{\cos (\pi k(N-1) / N)} \right).
\label{eq1}
\end{equation}

By performing trigonometric transformations, we obtain:
\begin{equation}
\varphi_Y[k]|_{k \neq 1} = \Delta \varphi_Y[k]|_{k \neq 1} =  -\frac{\pi} {2}-\frac{\pi k(N-1)} {N)} .
\label{eq1}
\end{equation}

Please note that Equation (4.16) is obtained under the condition $\alpha$ = 0 and $m=1$. Figure 3 depicts the dependencies of the measurement errors of the amplitude and phase spectra.

The dependencies were obtained through simulation using the analytical expressions (4.9) for the amplitude spectrum and (4.14), (4.16) for the phase spectrum. The graphs correspond to the derived analytical dependencies (4.9), (4.14), and the results of the simulation assuming that the ADC quantization error can be described using equation (4.4). The parameters of simulation match those adopted for constructing the graphs in Figures 1.

\section{Conclusion}

In this study, the influence of quantization noise on the spectral characteristics of the signal and the estimation of measurement errors in spectrum analysis caused by ADC quantization noise was investigated. It was confirmed through literature sources that quantization noise leads to distortion of the spectrum of the digital signal at the output of the ADC. To achieve more accurate and reliable spectrum measurements, an error estimation was performed to consider the impact of quantization noise on the spectral data. This is important for obtaining more accurate results and ensuring high-quality measurements of spectral characteristics. Additionally, it enables improved accuracy and reliability of measurement results and facilitates a more detailed analysis of signal spectral characteristics. 
Furthermore, further research can contribute to the development of methods for estimating measurement errors in spectrum analysis, considering other potential sources of measurement errors, as well as the development of compensation algorithms to reduce the impact of quantization noise.

\begin{Backmatter}

%\paragraph{Acknowledgments}
%We are grateful for the technical assistance of A. Author.

%\paragraph{Funding Statement}
%This research was supported by grants from the <funder-name><doi>(<award ID>); <funder-name><doi>(<award ID>).

\paragraph{Competing Interests}
None

%\paragraph{Data Availability Statement}
%A statement about how to access data, code and other materials allowing users to understand, verify and replicate findings --- e.g. Replication data and code can be found in Harvard Dataverse: \verb+\url{https://doi.org/link}+.

\paragraph{Ethical Standards}
The research meets all ethical guidelines, including adherence to the legal requirements of the study country.

\paragraph{Author Contributions} All authors approved the final submitted draft.
%Please provide an author contributions statement using the CRediT taxonomy roles as a guide {\verb+\url{https://www.casrai.org/credit.html}+}. Conceptualization: A.A; A.B. Methodology: A.A; A.B. Data curation: A.C. Data visualisation: A.C. Writing original draft: A.A; A.B. 

% For JDM please remove this \begin{thebibliography}...\end{thebibliography} list.
% Use biblatex-apa (see instructions in preamble) instead, and write \printbibliography here to print the reference list in APA7 style.

\end{Backmatter}


\begin{thebibliography}{}

\bibitem[Volodarskyi Ye.T. et al.(2020)]{bib1}
\textbf{Volodarskyi Ye.T.; Dobroliubova M.V.; Kosheva L.O.} (2020) Informatsiyno-vymiriuvalni systemy ta nevyznachenist. (in Ukraine) \textit{Ukrainskyi metrolohichnyi zhurnal/Ukrainian Metrological Journal}, 3A, {30-35}.

\bibitem[WIDROW, Bernard et al.(1996)]{bib2}
\textbf{WIDROW, Bernard; KOLLAR, Istvan; LIU, Ming-Chang.} (1996) Statistical theory of quantization. \textit{IEEE Transactions on instrumentation and measurement}, 45.2, {353-361}.

\bibitem[Kester, Walt(2004)]{bib3}
\textbf{Kester, Walt.} (2004) Analog-digital conversion. Norwood, MA: Analog Devices.

\bibitem[Kester, W.(2009)]{bib4}
\textbf{Kester, W.} (2009) Taking the Mystery out of the Infamous Formula, "SNR = 6.02N + 1.76dB," and Why You Should Care. \textit{MT-001}. Norwood, Analog Devices.

\bibitem[Kester, W.(2009)]{bib5}
\textbf{Kester, W.} (2009) Understand SINAD, ENOB, SNR, THD, THD + N, and SFDR so You Don't Get Lost in the Noise Floor. \textit{MT-003}. Norwood, Analog Devices.

\bibitem[Cruz Serra A. et al.(2004)]{bib6}
\textbf{Cruz Serra A., Da Silva M.F., Ramos P., Michaeli L.} (2004) Fast ADC testing by spectral and histogram analysis. \textit{Proceedings of the 21st IEEE Instrumentation and Measurement Technology Conference (IMTC 04)}, {823-828}.

\bibitem[Bellan D. et al.(1999)]{bib7}
\textbf{Bellan D., Brandolini A., Gandelli A.} (1999) Quantization theory a deterministic approach. \textit{IEEE Transactions on Instrumentation and Measurement}, 48(1), {18-25}.

\bibitem[Bellan D. et al.(1998)]{bib8}
\textbf{Bellan D., Brandolini A., Gandelli A.} (1998) ADC nonlinearities and harmonic distortion in FFT test. \textit{IEEE Instrumentation and Measurement Technology Conference}, IMTC/98, Vol. 2, {1233-1238}.

\bibitem[Brandolini A. et al.(1992)]{bib9}
\textbf{Brandolini A., Gandelli A.} (1992) Testing methodologies for analog-to-digital converters. \textit{IEEE Transactions on Instrumentation and Measurement}, 41(5), {595-603}.

\bibitem[Melnychuk, V.M. and Polikarovs'kykh, O.I.(2017)]{bib10}
\textbf{Melnychuk, V.M., Polikarovs'kykh, O.I.} (2017) Analiz parametrov tsyfro-analogovoho peretvorennia u priamykh tsyfrovykh syntezatorakh chastoty (DDS).(in Ukraine) \textit{Visnyk Khmelnytskoho natsionalnoho universytetu. Tekhnichni nauky}, No. 6, {152-158}.

\bibitem[Michaeli L. et al.(2008)]{bib11}
\textbf{Michaeli L., Michalko P., Šaliga J.} (2008) Unified ADC nonlinearity error model for SAR ADC. \textit{Measurement}, 41(2), {198-204}.

\bibitem[Bellan D. et al.(1998)]{bib12}
\textbf{Bellan D., Brandolini A., Gandelli A.} (1998) Effects of ADC nonlinearities in sinewave amplitude measurement. \textit{IEEE International Conference on Electronics, Circuits and Systems}, Vol. 3, {449-452}.

\bibitem[Bellan D. et al.(1995)]{bib13}
\textbf{Bellan D., Brandolini A., Gandelli A.} (1995) Quantization theory in electrical and electronic measurements. \textit{Proceedings Integrating Intelligent Instrumentation and Control (IEEE Instrumentation and Measurement Technology Conference, IMTC/95)}, {494–499}.

\bibitem[Emanuel A.E.(2010)]{bib14}
\textbf{Emanuel A.E.} (2010) Power definitions and the physical mechanism of power flow. \textit{Wiley Chichester}.

\bibitem[Gherasim C. and Van den Keybus J.(2004)]{bib15}
\textbf{Gherasim C., Van den Keybus J.} (2004) DSP implementation of power measurements according to IEEE trial-use standard. \textit{IEEE transactions on instrumentation and measurement}, 53(4), {1086-1092}.

\bibitem[IFEACHOR, Emmanuel C. and JERVIS, Barrie W.(2002)]{bib16}
\textbf{IFEACHOR, Emmanuel C.; JERVIS, Barrie W.} (2002) Digital signal processing: a practical approach. \textit{Pearson Education}.

\bibitem[SEROV, Andrey N. et al.(2018)]{bib17}
\textbf{SEROV, Andrey N.; SEROV, Nikolay A.; MAKARYCHEV, Petr K.} (2018) Evaluation of the Effect of Nonlinearity of the Successive Approximation ADC to the Measurement Error of RMS. In: \textit{2018 International Symposium on Industrial Electronics (INDEL)}. IEEE, {1-6}.

\end{thebibliography}
\end{document}